\documentclass[twocolumn,showpacs,preprintnumbers,amsmath,amssymb,superscriptaddress]{revtex4}

\usepackage{graphicx}
\usepackage{dcolumn}
\usepackage{bm}
\usepackage{longtable}

\begin{document}


\title{Configurations and Observables in an Ising Model with Heat Flow}

\author{E. Agliari}
\affiliation{Dipartimento di Fisica, Universit\`a degli Studi di
Parma, viale Usberti 7/A, 43100 Parma, Italy}
\author{M. Casartelli}
\affiliation{Dipartimento di Fisica, Universit\`a degli Studi di
Parma, viale Usberti 7/A, 43100 Parma, Italy}
\affiliation{CNR-INFM, Gruppo Collegato di Parma, viale Usberti
7/A, 43100 Parma, Italy} \affiliation{INFN, Gruppo Collegato di
Parma, viale Usberti 7/A, 43100 Parma, Italy}
\author{A. Vezzani}
\affiliation{Dipartimento di Fisica, Universit\`a degli Studi di
Parma, viale Usberti 7/A, 43100 Parma, Italy}
\affiliation{CNR-INFM, Gruppo Collegato di Parma, viale Usberti
7/A, 43100 Parma, Italy}

\date{\today}

\begin{abstract}We study a  two dimensional Ising model between thermostats
at different temperatures. By applying the recently introduced KQ
dynamics, we show that the system reaches a steady state with
coexisting phases transversal to the heat flow. The relevance of
such complex states on thermodynamic or geometrical observables is
investigated. In particular, we study energy, magnetization and
metric properties of interfaces and clusters which, in principle,
are sensitive to local features of configurations. With respect to
equilibrium states, the presence of the heat flow amplifies the
fluctuations of both thermodynamic and geometrical observables in
a domain around the critical energy. The dependence of this
phenomenon on various parameters (size, thermal gradient,
interaction) is discussed also with reference to other possible
diffusive models.
\end{abstract}


\maketitle

\section{Introduction}
\label{intro}

The study of systems undergoing heat flows is a classical topic in
non equilibrium statistical mechanics. Several important results
have been obtained, especially for one dimensional models with
continuous symmetries, such as chains of anharmonic oscillators
(see e.g. \cite{livi} for a review). On the contrary, there are
very few results for discrete models in two dimensions. A
ferromagnetic rectangular Ising lattice with a ``cylindrical''
geometry, i.e. opposite borders at temperatures $T_1$ and $T_2$ in
one direction, and periodic conditions in the other one, has been
introduced in \cite{harris} by Harris and Grant, and in \cite{saito} by
Saito, Takesue and Miyashita (see also \cite{eisler} for recent developments
on related matter). However, severe restrictions on the admitted temperature
intervals were present in both papers, due to intrinsic limitations of the
microcanonical dynamics used there (Creutz or Q2R rules).

Such restrictions have been removed in \cite{cmv} by introducing a
peculiar new  dynamics, briefly denoted as ``KQ dynamics'',
combining the advantages of the Q2R and Kadanoff-Swift rules. In
this way, due to an effective ergodicity in the whole range of
temperature intervals $(T_1,T_2) $, steady states take place for
all imposed temperatures. In particular,  for $T_1
< T_c <T_2$ (where $ T_c$ denotes the equilibrium critical
temperature), different phases steadily coexist: a magnetized
phase near the cold border at $T_1$, a paramagnetic phase near the
opposite hot border at $T_2$, and an intermediate phase around the
region at energy density $E_c$, the mean energy corresponding, at
equilibrium, to $T_c$. Moreover, the transport properties of the
system are well described by introducing an energy dependent
diffusivity. This occurs in a smooth way, possibly except around
$E_c$, where the specific heat diverges and the diffusivity
vanishes in the thermodynamic limit.

Transport apart, an open problem - and our main item indeed - is
the physical relevance of such steady states, characterized by many
coexisting phases, as they are distinct from homogeneous
equilibrium states. More precisely, for local physical
observables, we ask if a portion of the cylinder has recognizable
and peculiar properties when a heat flow passes through it. In
particular, we shall concentrate on sections perpendicular to the
flow (columns, vertical bands). Two kinds of physical observables
will be considered: thermodynamic quantities, such as energy
density and magnetization, and geometrical-dynamical observables,
for which the role of the configurations driven by the dynamics is
predominant. The latter observables are based on the metric
properties of the configurations, which may involve very different
items: the integral of pointwise differences (i.e. the well known
Hamming distance), which in some cases assumes an ``energetic''
meaning, or the measure of differences in cluster distributions
(Rohlin distance), an information-based metrics requiring the
formalism of partition spaces.

The main point is the existence of an energy band ${\Delta E}$,
starting just below $E_c$, where the observable fluctuations  are
remarkably wider for a system undergoing a heat flow with respect
to thermalized or close systems. The same happens to the
distances between configurations. All this may be read as evidence
of a larger variability of the system when it is far from
equilibrium. These
features strongly depend on the size $L$, and they disappear in
the thermodynamic limit $L\to \infty$. More precisely, as
expected, they vanish as soon as the energy gradient between
neighbouring columns becomes infinitesimal and local equilibrium
is reached. However, since real systems are characterized by
finite gradients and finite sizes, such large fluctuations could
be relevant in the study of mesoscopic systems with stationary
flows.

We recall that there are examples of exotic
dynamics where the local equilibrium is not reached even for
infinitesimal gradients \cite{dhar}. Remarkably, also in such cases
fluctuations are larger in the presence of heat flow.

A number of questions arise. For instance, how much do these
features depend on the chosen dynamics? And which is the role of
the specific spin interaction? As for the former question,
the robustness of our results has been tested by many checks, improving, in
addition, the reliability of the results described in \cite{cmv}.
The latter question is evidently crucial for the possible physical
relevance of the results. Now, for a purely diffusive process,
e.g. a Random Walk (RW), analogous experiments clearly indicate
the absence of the described phenomena, showing the essential role
played by the interaction. However, a deeper insight on the nature
of admissible interactions would require a more sophisticated
analysis, not developed here. The same holds for the role of other
possible relevant parameters, such as the topology of the
underlying structure or the presence of noise in the interactions.

The paper is organized as follows: in Section \ref{sec:2} the
model is introduced, with KQ dynamics (\ref{model}), and with
definitions and notations for the quantities involved in
experiments (\ref{thermobs}, \ref{partitions}); in Section
\ref{sec:3} we review the main results obtained from  numerical
experiments. Problems recalled above (relevance of KQ dynamics on
the results, etc.) are discussed  in Section \ref{conc}, with
further comments and perspectives on future work. Finally, in the
Appendix, we summarize the essential information on the formalism
necessary to define the Rohlin distance in partitions spaces.

\section{Model, Dynamics, Notations}
\label{sec:2}
\subsection{The Cylindrical Ising Model}
\label{model} The cylindrical Ising model considered in
\cite{saito} and \cite{cmv} is a $L_X\times L_Y$ rectangular
lattice, with periodic conditions in the $Y$ direction  and open
boundaries in the $X$ direction. We assume $ L_X=L_Y =L$. The spin
variable $\sigma_{x,y}$ may be $1$ or $-1$, and adjacent opposite
spins give an energy unit to the system. Thus, by denoting
$\langle x,y \rangle $ the nearest neighbours of $(x,y)$, the
normalized total energy $E_{tot}$ is:
\begin{equation}
\label{entot} E_{tot} ~=~{1\over{4L^2}}\sum_{x,y}\sum_{\langle x,y
\rangle}{{1- \sigma_{x,y}\sigma_{\langle x,y \rangle}}\over{2}}~.
\end{equation}

The lattice is naturally sliced into ``columns'' with a circular
symmetry. The first and last columns, i.e. the left and right
borders, interact with two thermostats, simulated by two sets of
supplementary columns evolving with the usual equilibrium
Metropolis algorithm (see \cite{cmv} for details). The Boltzmann's
constant $K$ is assumed to be $1$.

Internal sites must evolve preserving the energy, and the microcanonical
rule used throughout the paper is the KQ dynamics introduced in \cite{cmv},
for the reasons discussed there. In order to define such a dynamics,
we must previously recall the Q2R and Kadanoff-Swift (KS) moves:

\noindent {\bf Q2R move:} in every chosen site the spin  is forced to flip
whenever energy is preserved, i.e. when half spins in the neighborhood
are up and half are down (see e.g. \cite{vich,pom,toff}).

\noindent {\bf KS move:} consider a diagonal with two opposite spins,
and exchange them whenever energy is preserved (see \cite{kadanoff}).

The second-neighbours exchange in KS is essential for the
dynamization of otherwise frozen configurations near the
cold border, ensuring an effective transitivity in the configuration
space. Then the evolution rule may be defined as follows:

\noindent {\bf KQ Dynamics:} a single KQ step is a sequence of
$L\times L$ randomly alternated Q2R and KS moves on randomly
chosen sites and diagonals. Such a step defines the natural time
unit $\tau $.

Besides tests already performed in \cite{cmv}, the reliability of
the KQ dynamics has been successfully checked by looking at the
robustness of the results with respect to various perturbations. A
meaningful test, for instance, consists in a neat change of the
randomness criterion in the choice of sites and diagonals to be
moved. By using a RW path (which could be also a physically
reasonable procedure) we obtained indeed the same results,
possibly apart the time scale. In all cases, a steady state is
easily established.

\begin{figure}
\resizebox{0.9\columnwidth}{!}{
\includegraphics{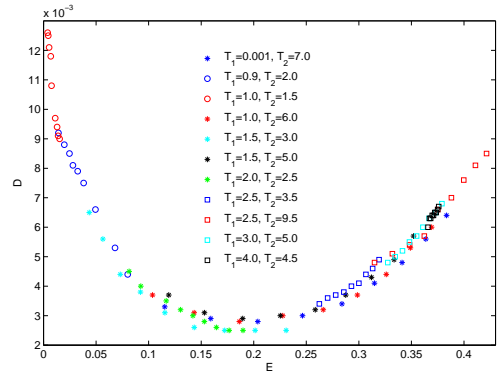}}
\caption{Diffusivity calculated as in reference \cite{cmv} for
$L=16$ and several $\Delta T$, proving the consistency of
assumptions also for small size (and high gradients) systems.}
\label{Fig_1}
\end{figure}

Another important aspect we have verified is that  even for small
systems ($L=16$) with large temperature differences
($T_1=0,~T_2=7$) the energy flow can be described by means of a
Fourier-like equation with an energy dependent diffusivity.
Therefore, data reported in Fig.~\ref{Fig_1} should be seen as an
improvement of those in the figure 10 of \cite{cmv}. This confirms
the correctness of the ansatz and the reliability of the results
presented there also very far from local equilibrium, i.e.
independently of any reference to quasi equilibrium local
temperature. Indeed, it is worth underlining that, in this
microcanonical context, and especially for small sizes, the local
temperature is not definite inside the lattice. Therefore, the
appropriate quantity characterizing local properties is the mean
local energy.

\subsection{Thermodynamic observables}
\label{thermobs} Typical quantities considered in  \cite{cmv} are
the mean energy densities of the columns, or $\langle E_x
\rangle$, where $x$ is the column label, and averages run for each
$x$ on both time and column sites. This may be seen as a
particular case of a general frame. By averaging at every time $t$
along the $Y$ direction only, $E_x \equiv E_x(t)$ is a discrete
time series; analogously for the squared magnetization $M^2_x(t)$
of the $x$-th column. All typical statistical features of time
series, first of all time averages and mean square deviations, may
be easily calculated. As usual, well stabilized values out of long
runs will be considered equivalent to the asymptotic ideal values
for all practical purposes.

An interesting point consists in the systematic comparison
between the Ising model with a heat flow (or IMF, for brevity)
and the closed Ising model (or CIM), i.e. the two dimensional
toroidal lattice whose energy (a constant of motion in this case)
will be fixed with suitable criteria. Alternatively, one can
compare the IMF and the thermalized Ising model (or TIM) where
the flow is zero because the borders are fixed at the same temperature.

More precisely, such comparisons require the following steps: 1 -
evaluate the mean energy  $E_{\tilde x}$ of a particular column
$\tilde x$ in the IMF; 2 - fix equal border temperatures for the
TIM or the total energy of the system for the CIM in such a way
that the average energy of any column in these systems is equal
to $E_{\tilde x}$; 3 - follow the time evolution of the systems
(IMF, TIM and CIM) in order to obtain three sequences of
decorraleted values for the different observables  (e.g. $E_{\tilde
x}$ and $M^2_{\tilde x}$); 4 - compute statistical properties of
the obtained time series.

These comparisons aim to stress the influence (if any) of the
local flow on physical observables with respect to
different types of thermalized systems.

Of course, an additional check is the comparison between TIM and
CIM, which should converge to the same behaviour for all
observables at least when $L \to \infty$.

\subsection{Geometrical observables}
\label{partitions} In order to give evidence to possible
correlations between heat flow and configurational features, we
need  a different kind of observables. Such observables have
already been used to study equilibrium states in spin systems
(precisely Ising systems, with or without long range correlations)
proving useful in focusing certain peculiarities of configurations
around the critical phase \cite{parti2,entro}.

The precise definition of these quantities requires the formalism
of configuration and partition spaces, as briefly summarized  in
the Appendix. However, the main idea is the following: consider
the configuration of a column ${\bf a} \equiv {\bf a}(x,t)$ as a
discrete periodic array  of $L$ binary values. A probability
measure $\mu$ is easily defined on the array subsets by the
normalized number of nodes in each subset. This way, an array
${\bf a}$ (or more precisely the triple constituted by ${\bf a} $,
$\mu$ and the algebra of subsets) becomes a particularly simple
example of finite probability space. An array may be partitioned
into homogeneous clusters $\{ A_1, A_2, ...,A_n \}$ of
consecutively aligned spins, and this collection $\alpha \equiv \{
A_1, A_2, ...,A_n \}$ may be seen as an element of the ``partition
space'' $\mathcal{Z}_L$ built on the probability space ${\bf
a}(x)$. We have established a correspondence $\Phi$ between
configurations and finite measurable partitions, or, more explicitely
$$ \alpha(x,t) = \Phi ({\bf a}(x,t))~.$$
In this case, the natural order of the  cluster sequence
identifies the partition by the first $Y$ coordinates of each
cluster. Shannon entropy, conditional entropy, Rohlin and Hamming
distances between two arbitrary columns are therefore well defined
functionals (see Appendix). We shall consider in particular the
following observables:
\begin{enumerate}
\item the Rohlin distance at a time $t$ between partitions
$\alpha(x)$ and $\alpha(x+1)$ associated to consecutive columns
${\bf a}(x) $ and ${\bf a}(x+1)$ of the same system, i.e.
$$d_R(x,t)=d_R(\alpha(x,t),~\alpha(x+1,t)) ~;$$
this distance is a measure of the non similarity between adjacent columns,
with regard to the cluster distributions;

\item the Rohlin distance between decorrelated columns with the
same energy (same label $x$). This is a measure of the non
similarity between independent columns. Decorrelated
configurations can be obtained considering either two distinct
systems evolving independently, or the same system and an
evolution time $\Delta t$ much larger than the decorrelation time.
Therefore, the fistances we consider are
$$d_R(\alpha(x,t),\beta(x,t) ) \quad {\rm{or}} \quad d_R(\alpha(x,t),~\alpha(x,t + \Delta t)) $$
respectively.
\item the Hamming distance between two adjacent columns, as in item 1, i.e.:
$$d_H(x,t)= d_H ( {\bf a}(x,t), ~{\bf a}(x+1,t))~ ;$$
this is another and very different measure of non similarity, focusing
on pointwise differences, independently of the neighborhoods. Moreover,
in this case, $d_H(x,t)$ represents the energy between a column and the next one.
\end{enumerate}
The observables defined above, like the previous thermodynamic
quantities, produce discrete time series, admitting statistical
analysis (means, deviations, etc.).

\section{Numerical experiments}
\label{sec:3}

As anticipated, numerical experiments tend to stress the influence
of the heat flow on significant observables, by comparing IMF and
CIM or TIM. Data, in the following,  will refer to both  time
averages  and averages over multiple experiments. Time averages
extend as usual up to stable results. Actually, averages run
over $10^5~-~10^6$ sampled values, ensuring an excellent stabilization,
as if, for all practical purposes, the limit $ t \to \infty $ had been
reached.

\subsection{Energy and Magnetization}
\label{enemag} 
\begin{figure}
\resizebox{0.9\columnwidth}{!}{
\includegraphics{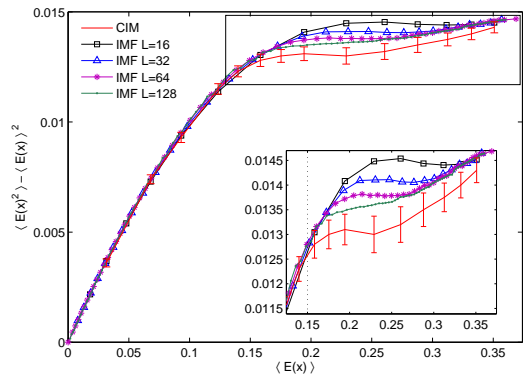}}
\caption{Energy fluctuations of columns vs. their mean energy,
for CIM of size $L=16$ (continuous line) and for IMF of different
sizes (as shown in the legend). The enlarged window shows details
in the region $E_c-\delta_1 , E_c+\delta_2 $.
}
\label{Fig_2}
\end{figure}

The first quantity we shall consider is the energy
density along the X direction, or $E_x$, $x$ being the label of
the array ${\mathbf a}(x)$, the configuration of the $x$-th
column. For each column the mean in the Y direction is always
assumed. Consider a system sampled at times $t_0, t_1, t_2,...$,
where the starting $t_0$ occurs after a suitable transient (e.g.
50 to 100 times $\tau $ for $ L=16$). Moreover, in order to have
sufficiently decorrelated configurations, $\Delta t \equiv
t_k-t_{k-1} > 100\tau$. Several $\Delta t$ have been tested. The
resulting time series $ \{ E_x(t_k )\}$  depends also on $L$ and
the border temperatures $(T_1,T_2)$. Then, for every $x$ there is
a mean energy density $\langle E_x \rangle$, and a Mean Square Deviation
$ F = \langle E_x^2 \rangle - \langle E_x
\rangle ^2$ (here $F$ stands for fluctuation). Such diagrams are
plotted in Fig.\ref{Fig_2} for $L=16,32, 64$ and $128$  at fixed
$(T_1,T_2) $. Here $T_1=0.01, ~T_2=4$, and the same in the
following, otherwise differently stated. In the same figure, at
the prescribed energies $\langle E_x \rangle $, the fluctuations of the
closed system (CIM), are plotted. Since they almost coincide for
different sizes, only the case $L=16$ is reported, with the error
bars. These diagrams show that:
\begin{itemize}
\item discrepancies $\Delta F$  between IMF and CIM, defined as
\begin{equation}
\Delta F= F_{_{IMF}} - F_{_{CIM}}
\end{equation}
(obviously, this definition may be adapted to various cases and observables),
 are especially important around the critical energy density $E_c $,
 in a range $\Delta E \equiv  (E_c-\delta_1, E_c+\delta_2) $, with $\delta_1$ very small.
 Moreover, $\Delta F > 0$, i.e. fluctuations are always greater for IMF.

\item Both the width of $\Delta E$ and the maximal amplitude of $
\Delta F$ depend on $L$. Indeed, as $L$ grows, $\Delta E$
decreases  and discrepancies $\Delta F$ {\it slowly} shrink. The
way $ \Delta E $ decreases seems faster in fact  than the
correlated way the $\max \mid \Delta F  \mid  $  vanishes.

\item By comparing data relevant to different sizes, we find that,
within $\Delta E$ and sufficiently far from $E_c$, $\Delta F$
scales like $1/L$. As for the very critical point $E_c$, our
numerical data do not allow any accurate prediction about the
behaviour of $\Delta F$, however, they suggest that $\Delta F$
decreases slower than $1/L$, as $L$ grows. Interestingly,
this can be read as a weak trace of criticality
around $E_c$.
\end{itemize}
We remark that the neighborhood of a certain column undergoing a
heat flow becomes more and more indistinguishable from an
equilibrium neighborhood as $L \to \infty$. Accordingly, it is
plausible that the column properties, inasmuch as they are related
to the state of its neighborhood, tend to mimic the equilibrium
properties in this limit.

\begin{figure}

\resizebox{0.9\columnwidth}{!}{
\includegraphics{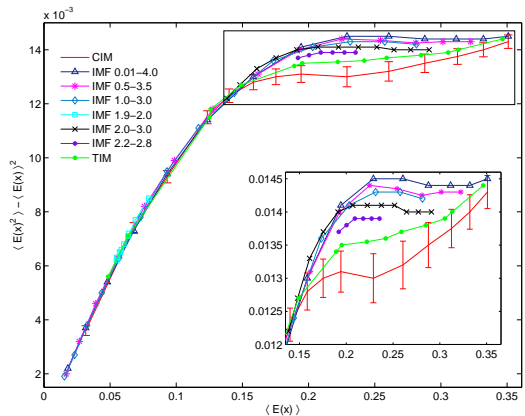}}
\caption{Energy fluctuations of columns vs. their mean energy, for
different $\Delta T$ in IMF, and for CIM (continuous line). The
enlarged window shows details putting in evidence the intermediate
TIM behaviour. For all systems the size is the same
$L=16$.
}
\label{Fig_3}
\end{figure}

In the same spirit, in Fig~.\ref{Fig_3} we can observe, at fixed
$L=16$, the effect of lowering the difference $\Delta T \equiv
T_2-T_1$ for IMF. The convergence of IMF to CIM is again clear,
starting from $E_c -\delta_1$ up to $E_c+\delta_2$, where
$\delta_1 $ is very thin and $\delta_2$ is smaller and smaller as
$L$ grows.

Neatly below $E_c-\delta_1$, or above $E_c+\delta_2$, the
coincidence between IMF and CIM is quite good for all $L$ and
$\Delta T$. A natural question is the reproduction of the same
results using a TIM instead of a CIM, i.e. a thermalized system
with equal border temperatures, such to give suitable mean
energies for comparisons. As a matter of fact, both CIM and TIM
give indeed qualitatively equivalent results with respect to IMF;
however, at the observed sizes, they do not coincide (see again
Fig.~\ref{Fig_3}). One expects, of course, that only for
sufficiently high $L$'s a good agreement will take place.

In general, the observed behaviour confirms a fact already noticed
in \cite{cmv}, i.e. enlarging $L$ is equivalent to zooming on
a system with a lower $\Delta T$, so that the
thermodynamic limit should give to every column the same features
of a system in local equilibrium. Clearly, such a zooming property
is not an absolute equivalence, since a finite size TIM cannot
reproduce an infinite size IMF. The equivalence refers only to the
onset of local equilibrium due to the vanishing of the gradient
between left and right side of each column. Moreover, critical
properties could disturb the continuity of this process around
$E_c$.

\begin{figure}
\resizebox{0.9\columnwidth}{!}{
\includegraphics{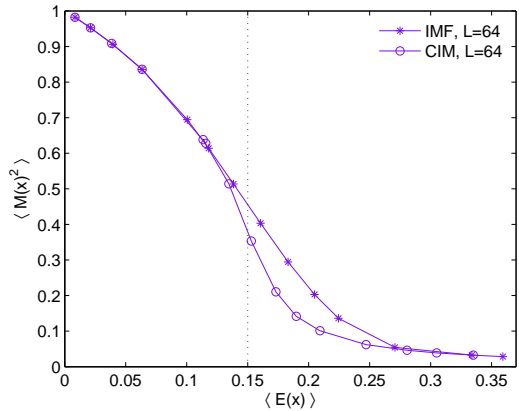}}
\caption{Mean values of squared magnetization $M^2$ of columns vs.
mean energy for IMF (stars) and CIM (circles). The right side of
the figure is equivalent to the $M$-fluctuations. } \label{Fig_4}
\end{figure}

Consider now the squared magnetization $M^2$, which above the
critical energy  coincides with the mean square deviation of $M$.
For a fixed size (here $L=64$), in Fig~.\ref{Fig_4} we plot the mean
values of $M^2$ vs. energy:  the IMF diagram is neatly above the
CIM diagram in the same region previously identified by energy
fluctuations, from $E_c - \delta_1 $ up to $E_c+\delta_2$. Hence,
in the same domain, also the magnetization fluctuations are larger
in the IMF system.

\subsection{Metric properties}
\label{metric} 
\begin{figure}
\resizebox{0.9\columnwidth}{!}{
\includegraphics{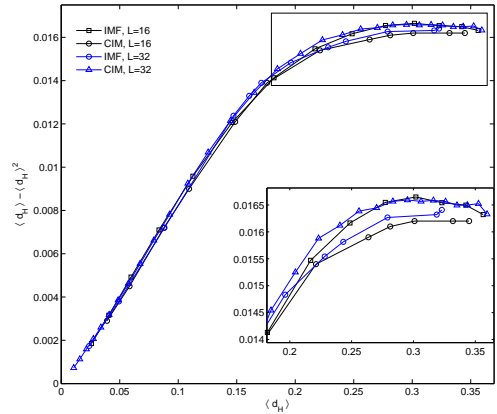}}
\caption{Hamming distance fluctuations of adjacent columns vs. the
distance self (or longitudinal energy) for several sizes of IMF,
showing the progressive convergence to CIM in the same region
previously indicated.
}
\label{Fig_5}
\end{figure}

The energy  between a column and the two adjacent
ones (X direction) should feel, in principle, the asymmetry
between left and right neighbourhoods. Clearly, as remarked in
subsection \ref{partitions}, such a longitudinal energy  between
close columns coincides with their Hamming distance $d_H$ (see
Appendix for definitions), giving this metric concept also a
physical interpretation. In Fig.~\ref{Fig_5}, the expected
difference between IMF and CIM for this quantity may be easily
recognized, once again in the same region previously evidenced by
energy density and magnetization.

\begin{figure}
\resizebox{0.9\columnwidth}{!}{
\includegraphics{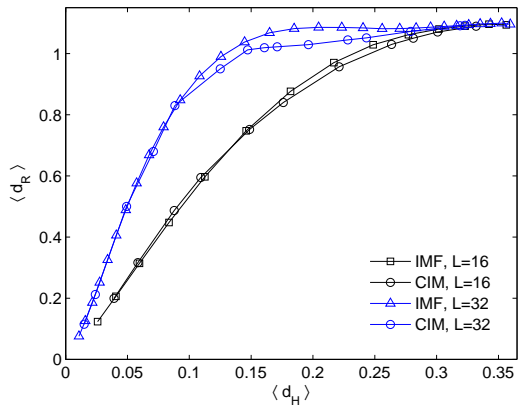}}
\caption{Rohlin distance between consecutive rows vs. Hamming
distance between the same rows. Results pertaining to IMF and CIM
are compared. } \label{Fig_6}
\end{figure}

\begin{figure}
\resizebox{0.9\columnwidth}{!}{
\includegraphics{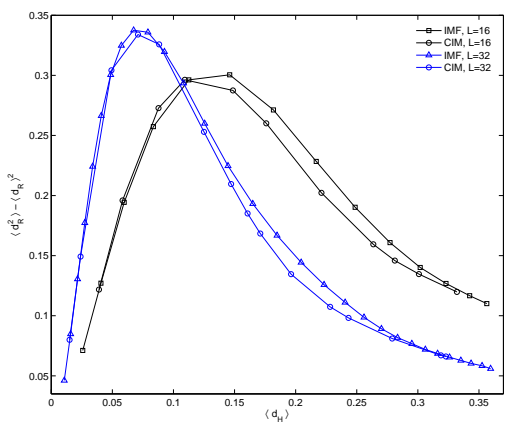}}
\caption{Rohlin distance fluctuations vs. the Hamming distance for
IMF and CIM and two sizes ($L=16,~32$). } \label{Fig_7}
\end{figure}

A quantity directly related to the configurations, more precisely
to the cluster distributions, is the Rohlin distance $d_R$ (see
Appendix), which may be measured with various attitudes. For
generic partitions, $d_R(\zeta ,\eta)$ is the amount of
information necessary to distinguish $\zeta$ from $\eta$, i.e. a
measure of their non-similarity. Such a non-similarity, in our
case, can regard both spatially or temporally distinct cluster
distributions. Since this appears deeply related to the
variability of configurations, $d_R$ is a good candidate, in
principle, to be an indicator of the influence of a gradient on
steady states. First of all, we consider couples of adjacent
columns, so that the longitudinal energy $d_H$ is a meaningful
alternative abscissa. The mean values and fluctuations of $d_R$
vs. $d_H$ are plotted in Fig.~\ref{Fig_6} and \ref{Fig_7}
respectively, confirming the larger variability of IMF system.

It would be also interesting to understand if it is possible to
distinguish systems with or without heat flow by looking at a
single column. To this end we consider the sequence of
uncorrelated configurations at times $t_1\dots t_k$, calling
$\alpha_k \equiv \alpha(x,t_k)$ the corresponding partition for
the $x$-th column (see Appendix for details). We calculate  the
numerical sequence of distances: $d_R(\alpha_k,\alpha _{k+1})$.
Such a sequence follows the ``novelty creation'' along an orbit
for every examined column, whereas the previous sequence followed
the evolution of an isochronous gradient of novelty between
adjacent columns. In both cases, fluctuations give overall
estimates of such dynamic or isochronous variability.

\begin{figure}
\resizebox{1.0\columnwidth}{!}{
\includegraphics{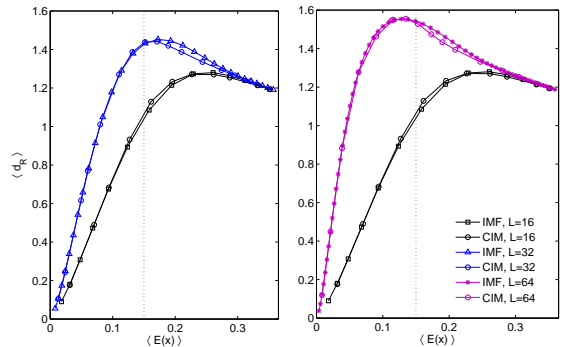}}
\caption{Time averaged Rohlin distances of single columns as a function of the
average energy of the column itself. Mean values for $L=16,32$
(left window) and $L=16,64$ (right window) are depicted, as shown
in the legend.
}
\label{Fig_8}
\end{figure}

In Fig.~\ref{Fig_8} we observe the behaviour of time averaged
$d_R(\alpha_k,\alpha _{k+1})$ for $L=16,32,64$ as a function of
the energy of the corresponding columns. For clearness, we have
splitted the comparison in two frames, 16-32 and 16-64
respectively. Apart the incidental inversion between IMF and CIM
at $L=16$, at larger $L$,  IMF-distances are greater than the
corresponding CIM-distances. Once again  the larger
variability of the system presenting heat flow is evidenced.

We note also that the maximum evolves with $L$: the peak grows
logarithmically, as expected, while the peak abscissa slowly decreases.

\begin{figure}
\resizebox{0.9\columnwidth}{!}{
\includegraphics{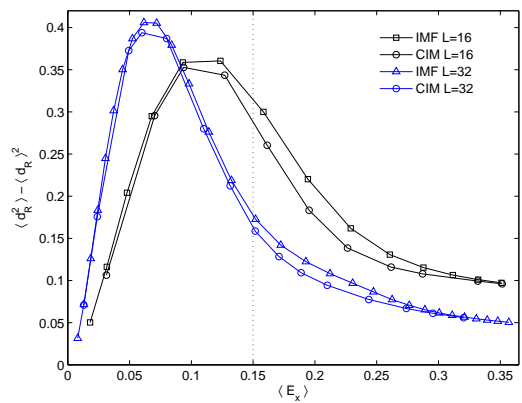}}
\caption{Fluctuations of Rohlin distance for single columns vs.
their mean energy for IMF and CIM and two sizes ($L=16,~32$).
}
\label{Fig_9}
\end{figure}

As to fluctuations, results summarized in Fig.~\ref{Fig_9} are
extremely similar to those in the previous Fig.~\ref{Fig_7}.

Three points have to be stressed:
\begin{itemize}
\item  the remarkable likeness between Figures \ref{Fig_7} and
\ref{Fig_9} is far from being trivial, since partitions are
strictly correlated in the former case, uncorrelated in the
latter; \item  fluctuations are only slightly wider in the
uncorrelated cases;
\item  the region interested by a discrepancy
between IMF and CIM is again the same, possibly with a small shift
toward low energies for the left bound.
\end{itemize}
A further remark is that nothing would be different in
Fig.~\ref{Fig_9} using partitions from two independent systems:
this confirms the complete decorrelation of configurations along
an orbit within $\Delta t$.

\section{Conclusions and perspectives }
\label{conc}

All numerical experiments on the Ising cylindrical model converge
on the fact that the imposed heat flow reveals in a wider
amplitude of fluctuations for local observables. Recalling the
robustness of present results with respect to small perturbations
of the dynamics, as remarked in (\ref{model}), a natural question
arises: how much does this behaviour depend on the IMF
peculiarities? In other terms, would an asymmety in the boundary conditions,
as the left-right temperature difference in our model,
be automatically translated into amplified fluctuations, when imposed
on a generic lattice system? If so,  being a general consequence
of spatial asymmetry in probabilistic processes, this feature
would be very weakly related to physics. We would argue, on the
contrary, that the observed behaviours of IMF vs. CIM or TIM  are
non trivially related to real mesoscopic features of a magnetic
system.

First of all, a point stressing the physical meaning of our
experiments is that the influence of heat flow on observables
appears to be deeply related to the peculiar way an Ising
rectangular model passes through the critical region. The
amplification, as remarked, does not regard indeed the whole of a
steady state, but only a relevant neighborhood of the magnetic
transition. On the contrary, for small values of $T_1$ and high
values of $T_2$, observables in the regions close to the borders
are practically indistinguishable from those in equivalent
equilibrium states. This last feature may be understood in terms
of typical configurations: near the cold border, there are indeed
sparse spots of one or two sites, making  the left and right
neighborhoods of the observed column practically identical. The
same happens near the hot border, provided that the temperature is
sufficiently high to establish a uniform disorder, this time
because of the irreducible fragmentation into thin clusters. Only
in the intermediate region there is a meaningful difference
between left and right sides, reflecting the growth and subsequent
fragmentation of clusters in the X direction. Columns are slices
of such clusters, with a shape dependence on $x$ heavily related
the properties of the Ising system.

A further indication that an asymmetry in boundary conditions is
not sufficient to explain the larger variability of IMF is
provided by a simple study of the paradigmatic model of non
interacting diffusion, i.e. RW on a lattice. Precisely, by
imposing different densities of walkers at the borders, it is
possible to show that, even in the presence of a strong density gradient,
fluctuations in the system remain unchanged.

Hence, a purely diffusive RW is too poor to reproduce the
behaviour we have observed in the Ising model, where evidently
interaction plays a fundamental role. In the same way, the very
existence of a critical temperature (or energy), which is
certainly related to the observed effects, is irreproducible by
simple RW. In order to clarify the subject, local interactions
should be introduced in the RW model, mimicking the role of the
energy dependent diffusivity in IMF. This may be done in several
ways, and studies in this direction are in progress, as well as
tests on totally different dynamical systems (e.g. asymmetric
sandpiles). All this will be fully reported in another paper.

Finally, we remark that the relevance of a finite (i.e. non
infinitesimal) thermal gradient, or the  consequent  vanishing
of $\Delta E$ and $\Delta F $ in the thermodynamic limit, do not
imply that the observed effects are physically meaningless. There
are no reasons indeed to consider finite size properties as
unphysical. A mesoscopic situation ($L$ finite) with peculiar
non-equilibrium features could be equally or even more interesting
from a physical point of view.

Acknowledgments

We thank N. Macellari and E. Vivo (Parma) for important discussions in the early phase of the work.

\appendix
\section{Appendix: Configurations, Partitions Spaces and Distances}
\label{appendix}

 Let $\mathbf{M}$ be a graph with $L$ nodes or sites ${a_j}$ assuming
 values in an alphabet
$\mathbf{K}$. A configuration on $\mathbf{M}$ is a whole set
${\mathbf{a}} =\{a_j\}, a_j \in \mathbf{K}$. It is an element of
${\mathcal{C}}= {\mathcal{C}(\mathbf{M})}$, the set of all $|
{\mathbf{K}}|^L$ possible states of the lattice. For instance, if
$\mathbf{M}$ is a discrete array (as in the case of our columns)
or a square lattice, and ${\mathbf{K}} = \{-1,1\}$,
 this description  fits Ising-like systems.

A \textit{path}, is a sequence of ``near'' sites, and a
\textit{connected} cluster is a set of sites with the same value
in $\mathbf{K}$ which are connected by a path. For general graphs,
 clusters are connected but not necessarily simply
connected sets. Since every site belongs
to a single cluster, clusters $A_{k}$ are disjoint subsets of
$\mathbf{M}$ and $ \bigcup_{k} A_{k} = \mathbf{M}$. In other
terms, the clusters collection is a ``finite partition'' of
$\mathbf{M}$, whose subsets $\{A_k\}$ constitute its ``atoms''.
The partition space $ \mathcal{Z} =  \mathcal{Z(\mathbf{M})}$ is
the set of all finite partitions of $\mathbf{M}$.
The correspondence $\Phi : \mathcal{C} \rightarrow \mathcal{Z}$
between a configuration $\mathbf{a} \in \mathcal{C}$ and the
clusters partition $\alpha\equiv (A_1,...,A_N) \in \mathcal{Z}$,
i.e. $\alpha = \Phi(\mathbf{a})$, is ``many to one'', because
configurations generated by permutations in $\mathbf{K}$ are
mapped into the same partition.

For every subset $A$ of $\mathbf{M}$, let  $~\mu(A) $ be the normalized number of nodes in
$A$. This defines a probability measure $\mu$ in
the algebra $\mathcal{M}$ of subsets of $\mathbf{M}$.

For standard operations on partitions in $\mathcal{Z}(\mathbf{M})$
classical textbooks are e.g. \cite{bill,AA,sinai,rohlin}. For
applications in the spirit of our demands, see also
\cite{parti2,entro,parti1,soc}. Here we only recall the
definitions of Shannon entropy and Rohlin distance.

Let $\alpha=(A_1,...,A_N)$ be a partition: its Shannon entropy $H
(\alpha)$ is
\begin{equation}
\label{shannon}
  H(\alpha)= -\sum_{i=1}^N \mu(A_i)\ln \mu(A_i) ~.
\end{equation}
The Shannon entropy does not depend on the shapes of the atoms,
but only on their measures. If  $\beta=(B_1,...,B_M)$ is another
partition, shapes implicitly influence the \textit{conditional}
entropy of $\alpha$ with respect to $\beta$:
\begin{equation}
  H(\alpha|\beta) = -\sum_{i=1}^N\sum_{k=1}^M \mu(A_i\cap B_k)\ln\frac{\mu(A_i\cap
  B_k)}{\mu(B_k)}~.
\end{equation}
Then, the Rohlin distance $d_R $ between partitions is defined by
\begin{equation}
\label{rohlin} d_R (\alpha,\beta)
=H(\alpha|\beta)+H(\beta|\alpha)~.
\end{equation}
This makes $\mathcal{Z}(\mathbf{M})$ a metric space. The Rohlin
distance expresses how different two partitions are.

If $\mathbf{K}$  itself is a metric space (e.g. a
numerical set with the usual distance between numbers),  one can
also consider  in ${\mathcal{C}(\mathbf{M})}$ the Hamming distance
$d_H$ which, for configurations $\mathbf{a}$ and $\mathbf{b}$, is
defined by the functional
\begin{equation}\label{hamming}
d_H ({\mathbf{a}},{\mathbf{b}}) =   \sum_j
|~b_j-a_j~|~
\end{equation}
(possibly normalized by dividing by $L$).
In our case, as noticed in Section \ref{metric}, the Hamming  distance between adjacent
columns is the energy between them.

In general, Hamming and Rohlin distances are not directly comparable.
The former is between
\textit{configurations}, and it is sensitive only to actual values
of corresponding nodes, not to their distribution or neighborhood,
 whereas the latter is between
\textit{partitions}, and therefore it is sensitive to the cluster
shapes. In principle, $d_R$ and $d_H$ may give very different
information. With a binary alphabet, for instance,
complementary configurations have maximal Hamming  distance ($d_H =
L$), while the corresponding partitions coincide ($d_R=0$).

 If a configuration $ {\mathbf{a}}\in \mathcal{C}$ has discrete evolution
 $$~{\mathbf{a}}, ~S{\mathbf{a}}, ~S^2 {\mathbf{a}},...,S^n{\mathbf{a}},...~$$
one can speak of ``configurations orbit''.
 The corresponding dynamics $\hat{S}$ on $ \mathcal{Z}$ is defined by
\begin{equation}
  \hat{S}\alpha=\hat{S}~ \Phi({\mathbf{a}} )= \Phi~(S{\mathbf{a}})
\end{equation}
so that to a configurations orbit there corresponds a partitions
orbit $\{  \hat{S}^n\alpha \} \equiv \{\Phi(S^n {\mathbf{a}}) \}$.
Clearly, the probability measure $\mu$
in $ \mathcal{Z}$  is not preserved by $ \hat{S}$,
because clusters do not evolve in themselves
but are redefined at every step by the pointwise dynamics in $\mathcal{C}$. However,
we are not interested here in indicators requiring a preserved measure,
such as Kolmogorov-Sinai entropy or Lyapunov
exponents.

Real valued observables $F$ or $\hat{F} $,
in $\mathcal{C}(\mathbf{M})$ or
$\mathcal{Z}(\mathbf{M})$, give rise to ``time series''
$\{f_k \} = \{F(S^k {\mathbf{a}}) \}$ or $\{  \hat{f}_k\} = \{ \hat{F}(\hat{S}^k \alpha)
\}$. Such time series are typical objects of our investigations.

This formalism applies in principle to every kind of lattices and
discrete dynamics. Note however that when $\mathbf{M}$ is a one dimensional array, as in the case considered here, the Rohlin distance is essentially simpler than in the two-dimensional case, because of the geometrical nature of the atoms contours: points in the former case, possibly cumbersome paths in the latter (see e.g. \cite{parti2,entro,soc}). For the Hamming distance, on the contrary, the computational complexity would be almost the same.

\end{document}